\documentclass{appolb}
\usepackage{epsfig}
\begin{document}
\title{Transverse momentum and multiplicity fluctuations in Ar+Sc collisions at the CERN SPS from NA61/SHINE%
\thanks{Presented at Critical Point and Onset of Deconfinement 2016, Wroclaw, Poland, May 30th - June 4th, 2016}%
}
\author{Evgeny Andronov for the NA61/SHINE Collaboration
\address{Saint Petersburg State University, ul. Ulyanovskaya 1, 198504, Petrodvorets, Saint Petersburg, Russia, {\tt e.v.andronov@spbu.ru}}
}
\maketitle
\begin{abstract}
The NA61/SHINE experiment aims to discover the critical point of strongly interacting matter and study the properties of the onset of deconfinement. For these goals a scan of the two dimensional phase diagram ($T$-$\mu_{B}$) is being performed at the SPS by measurements of hadron production in proton-proton, proton-nucleus and nucleus-nucleus interactions as a function of collision energy and system size. In this contribution preliminary results on transverse momentum and multiplicity fluctuations expressed in terms of strongly intensive quantities from the Ar+Sc energy scan will be presented. These fluctuations are expected to be sensitive to the existence of a critical point. The results are compared with results from the p+p and Be+Be energy scan as well as with NA49 measurements.
\end{abstract}
  
\section{Introduction}
The NA61/SHINE experiment~\cite{Abgrall:2014fa} is a multi-purpose fixed target experiment at the Super Proton Synchrotron (SPS) of the European Organization for Nuclear Research (CERN). The strong interaction programme of NA61/SHINE consists of studies of the onset of deconfinement (OD)~\cite{Gazdzicki:1998vd} in nucleus-nucleus collisions and search for the critical point (CP)~\cite{Fodor:2004nz} of strongly interacting matter.

NA61/SHINE performes measurements of hadron production in collisions of protons and various nuclei (p+p, Be+Be, Ar+Sc, Xe+La, Pb+Pb) at a range of beam momenta (13{\it A} - 158{\it A} GeV/c). It is expected that there will be a non-monotonic dependence of fluctuations of a number of observables in this scan due to the phase transition of strongly interacting matter and the possible existence of the CP~\cite{Stephanov:1999zu}. Some hints of this enhancement were already observed by the NA49 experiment~\cite{Grebieszkow:2009jr}.
\section{Fluctuation measures}
The volume of the system created in ultrarelativistic ion collisions fluctuates from event to event. In order to suppress contribution from this 'trivial' fluctuations strongly intensive observables are used \cite{Gorenstein:2011vq, Gazdzicki:2013ana}: 
\begin{eqnarray}
            &\Delta[A,B] = \frac{1}{C_{\Delta}} \biggl[ \langle B \rangle \omega[A] -
                        \langle A \rangle \omega[B] \biggr] \\
            &\Sigma[A,B] = \frac{1}{C_{\Sigma}} \biggl[ \langle B \rangle \omega[A] +
                        \langle A \rangle \omega[B] - 2 \bigl( \langle AB \rangle -
                        \langle A \rangle \langle B \rangle \bigr) \biggr],
\end{eqnarray}
where $\langle\cdots\rangle$ stands for averaging over all events and $\omega[X]$ is the scaled variance of any quantity $X$ defined as $(\langle X^{2}\rangle -{\langle X\rangle}^{2})/\langle X\rangle$. In case of joint transverse momentum $P_{T}$ and multiplicity $N$ fluctuations we define:
$A=P_{T}=\sum_{i=1}^{N}p_{T_{i}}$, $B=N$, $C_{\Delta}=C_{\Sigma}=\langle N\rangle \omega[p_{T}]$.

The advantage of strongly intensive quantities is that in the models of independent sources \cite{Bialas:1976ed} and in the models of independent particle production the contribution from volume fluctuations is eliminated, allowing us to probe the CP signals. Moreover, in the models of independent particle production $\Delta[P_{T},N]=\Sigma[P_{T},N]=1$. In recent measurements of NA61/SHINE \cite{Czopowicz:2015mfa} no anomaly attributable to the CP was observed neither in p+p nor in forward energy selected Be+Be collisions.
\section{Results}
The analysis was preformed for inelastic p+p interactions at 20, 30, 40, 80 and 158 GeV/c beam momenta, as well as Be+Be and Ar+Sc collisions at 19, 30, 40, 75 and 150{\it A} GeV/c selected for the smallest 5$\%$ of forward energies. Selection of the most central events by the forward energy was done using information from the NA61/SHINE forward calorimeter – the Particle Spectator Detector (PSD) \cite{Seryakov}.

In contrast to the previous measurements \cite{Czopowicz:2015mfa}, the analysis was restricted to the interval $0<y^{*}_{\pi}<y_{beam}$, where $y^{*}_{\pi}$ is the particle rapidity in the $c.m.s.$ under pion mass assumption. The lower cut is motivated by  poor azimuthal angle acceptance and stronger electron contamination at backward rapidities. The upper cut is used to reduce effects of diffractive collisions. Transverse momenta of all charged particles were restricted to $0<p_{T}<1.5$ GeV/c.

The results for p+p and Be+Be collisions were corrected for contributions from non-target interactions. For this purpose NA61/SHINE acquired data with both target inserted and removed. The contribution from non-target interactions was subtracted in the analysis. For Ar+Sc collisions these corrections were estimated to be negligible. The results were also corrected for detector inefficiencies and trigger biases by multidimensional corrections of the moments in p+p and Be+Be collisions and by separate moment corrections for Ar+Sc collisions.

The statistical uncertainties were estimated using the sub-sample method \cite{Czopowicz:2015mfa}, and the systematic uncertainties were estimated to be smaller than 10$\%$ for $\Delta[P_{T},N]$ and $\Sigma[P_{T},N]$. Further analysis of the systematic uncerainties will follow.
\subsection{Energy and system size dependence of $\Delta[P_{T},N]$ and $\Sigma[P_{T},N]$}
Figures 1 and 2 show the energy dependence of $\Delta[P_{T},N]$ and $\Sigma[P_{T},N]$ for p+p, Be+Be and Ar+Sc collisions for all charged, positively charged and negatively charged hadrons. The same results are presented in two-dimensional plots as unction of $\sqrt{s_{NN}}$ and mean number of wounded nucleons $\langle W\rangle$ (from Ref.~\cite{Broniowski:2007nz}) in Figs. 3 and 4. No dip-like or hill-like structures that could be related to the critical point of strongly interacting matter \cite{Vovchenko} are observed in these results. 
%uncomment the following lines to place a figure
\begin{figure}[htb]
\centerline{%
\includegraphics[width=4.5cm]{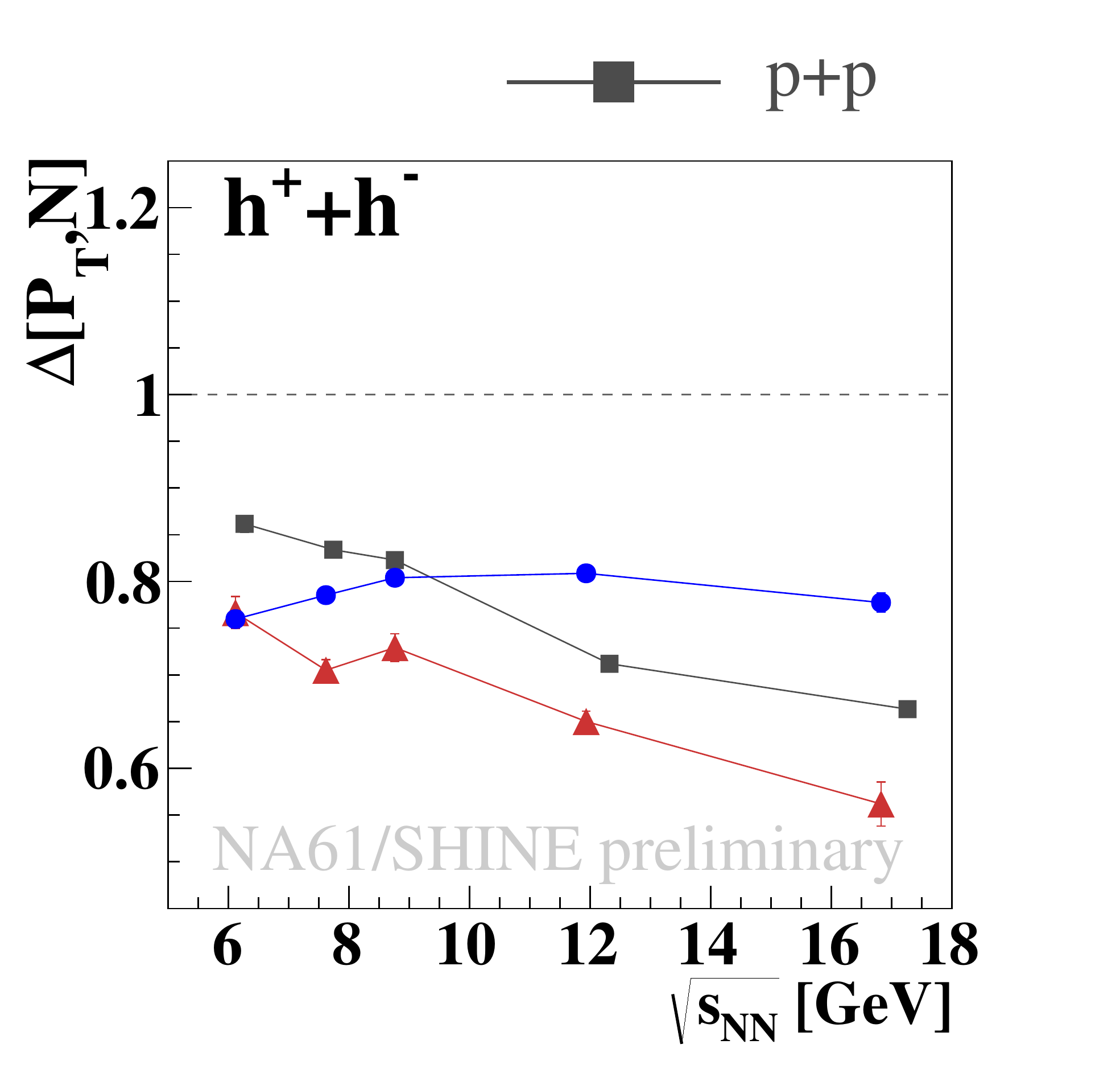}\includegraphics[width=4.5cm]{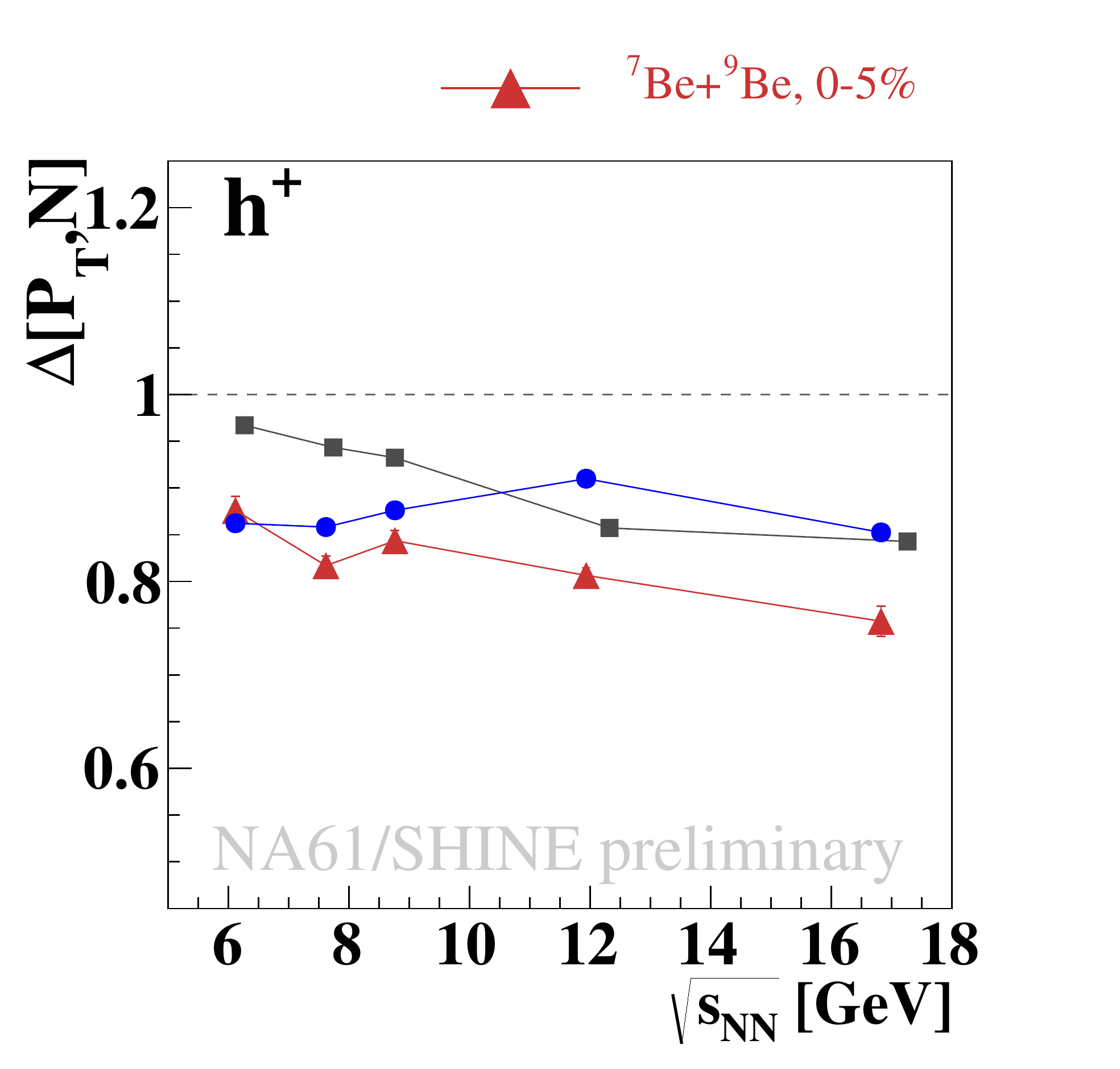}\includegraphics[width=4.5cm]{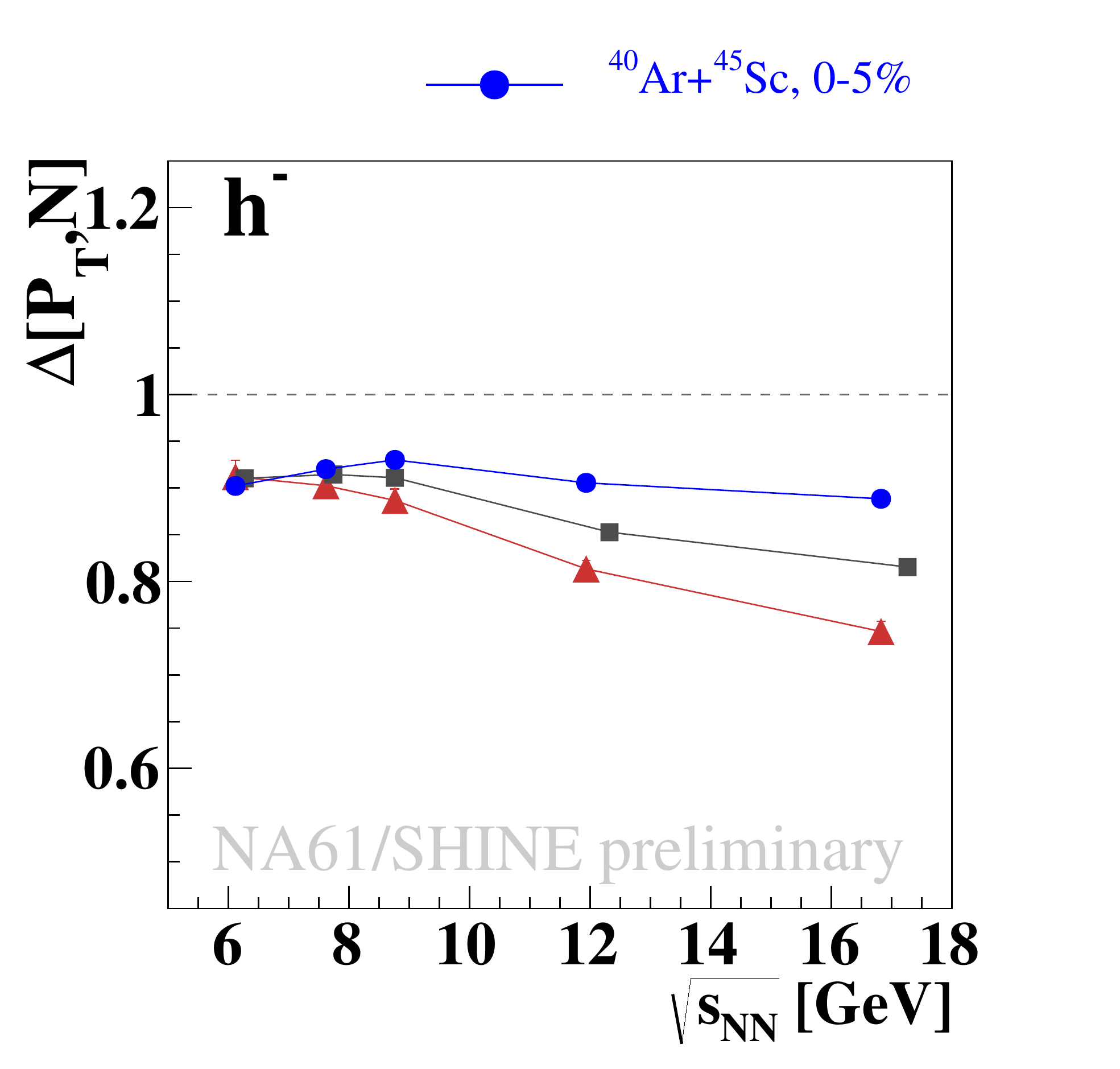}}
\caption{$\Delta[P_T, N]$ as a function of $\sqrt{s_{NN}}$ in inelastic p+p (grey squares), 0-5\% central Be+Be (red triangles) and Ar+Sc (blue circles) collisions at forward-rapidity, $0< y_{\pi}< y_{beam}$, and in $p_T<1.5$ GeV/c. Results are shown for all charged ($h^{+}+h^{-}$, left), positively charged ($h^{+}$, middle) and negatively charged hadrons ($h^{-}$, right).}
\label{Fig:F2HDeltaEn}
\end{figure}
\begin{figure}[htb]
\centerline{%
\includegraphics[width=4.5cm]{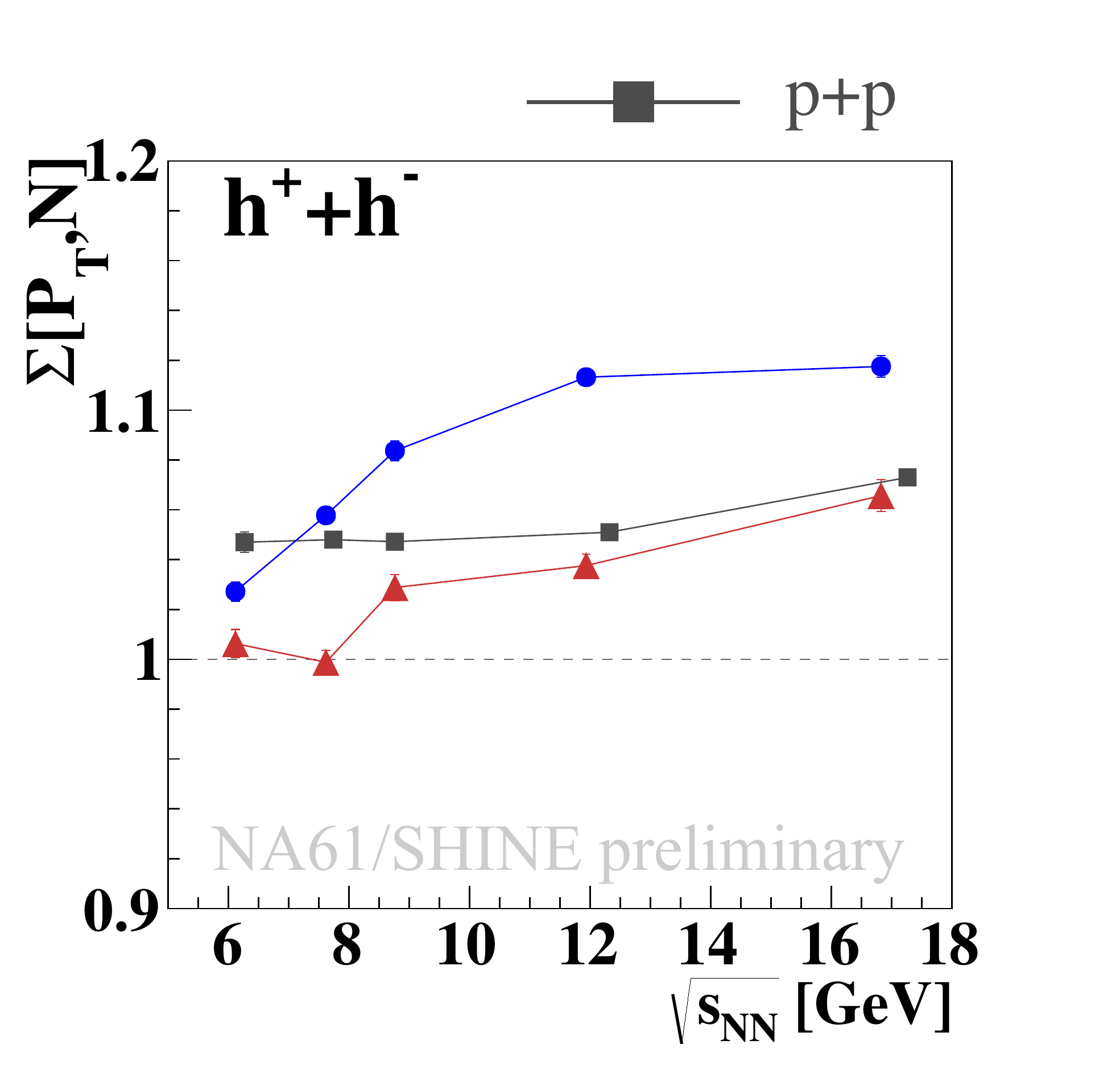}\includegraphics[width=4.5cm]{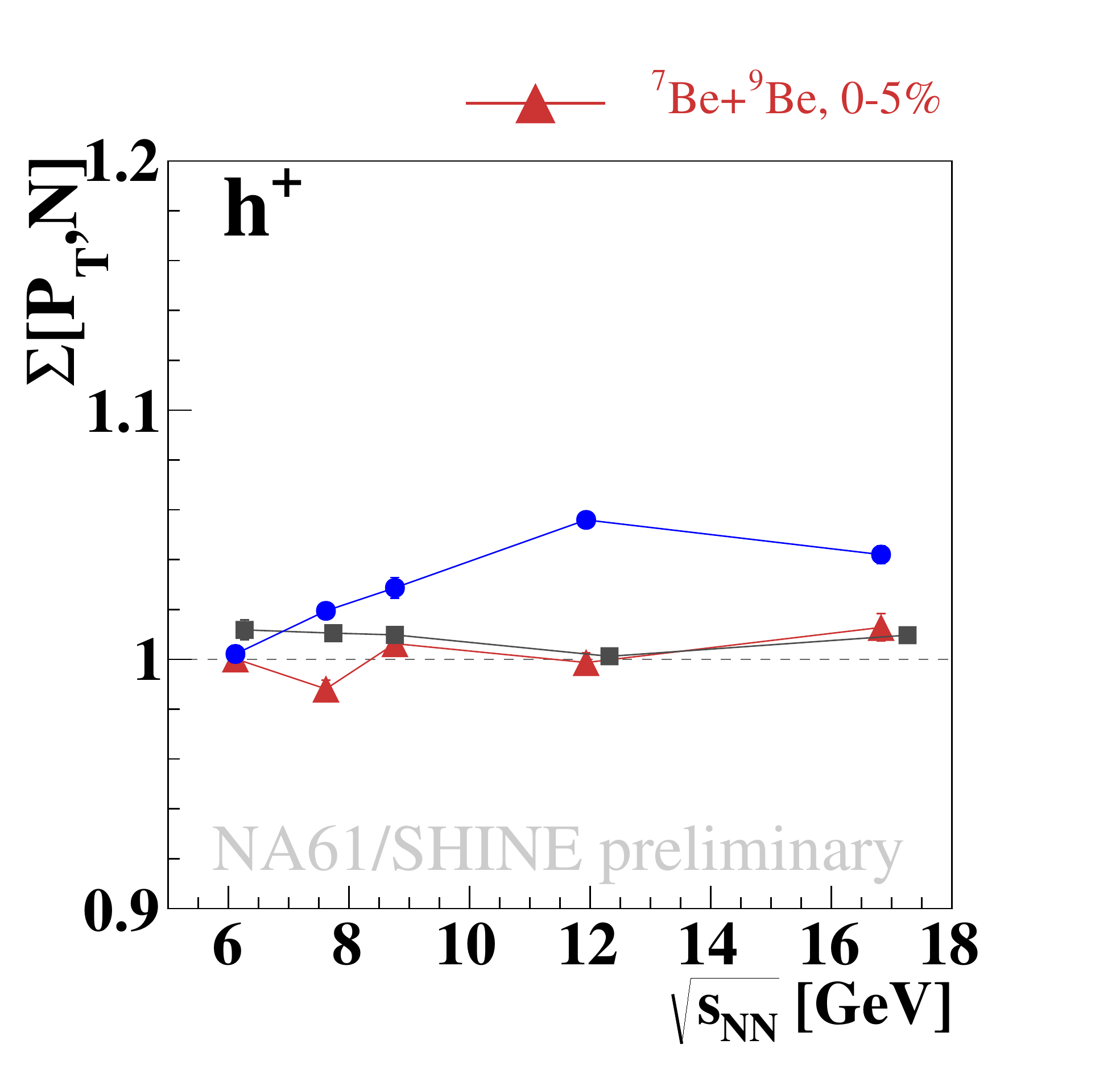}\includegraphics[width=4.5cm]{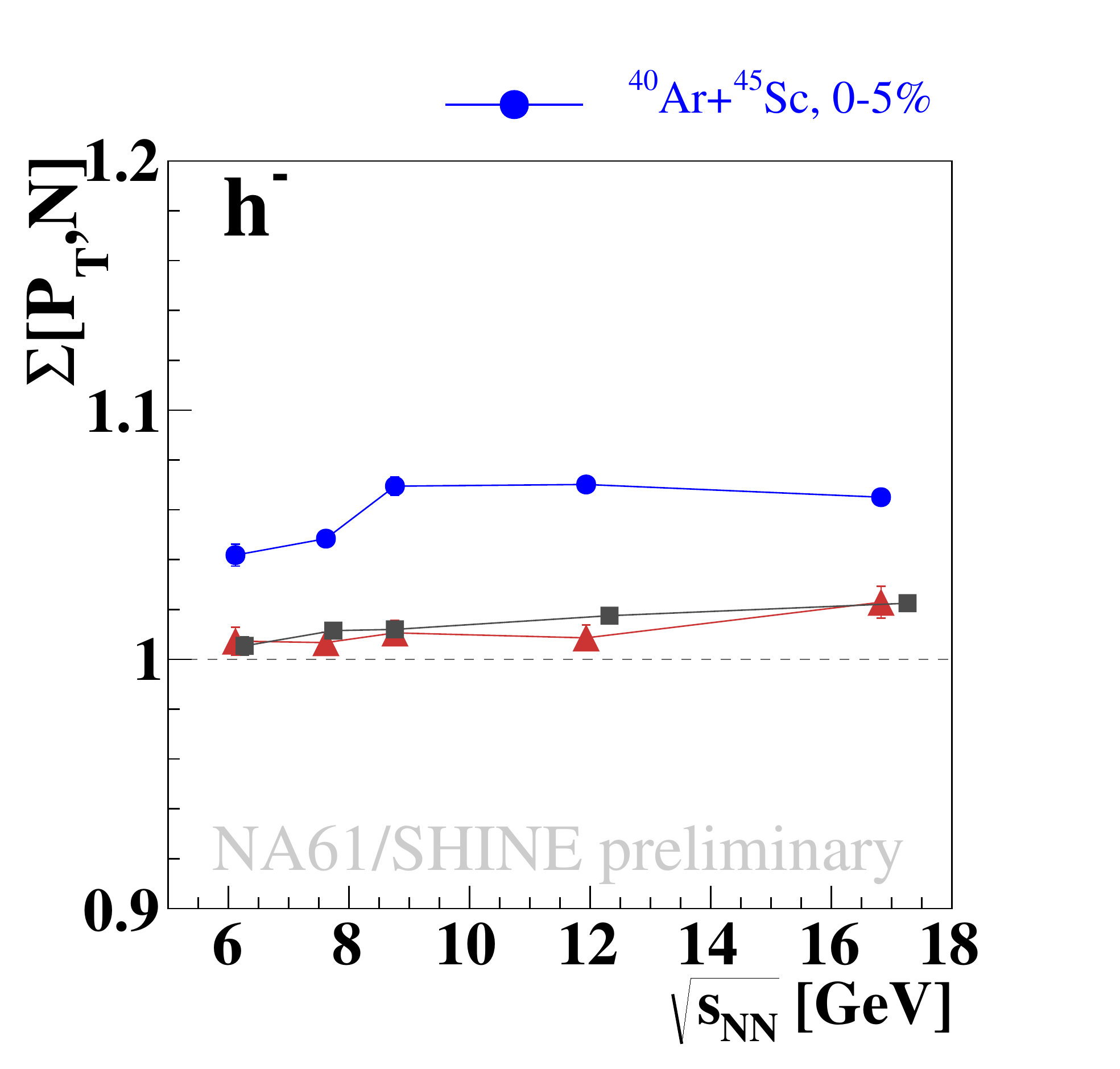}}
\caption{$\Sigma[P_T, N]$ as a function of $\sqrt{s_{NN}}$ in inelastic p+p (grey squares), 0-5\% central Be+Be (red triangles) and Ar+Sc (blue circles) collisions at forward-rapidity, $0< y_{\pi}< y_{beam}$, and in $p_T<1.5$ GeV/c. Results are shown for all charged ($h^{+}+h^{-}$, left), positively charged ($h^{+}$, middle) and negatively charged hadrons ($h^{-}$, right).}
\label{Fig:F2HSigmaEn}
\end{figure}
\begin{figure}[htb]
\centerline{%
\includegraphics[width=4.5cm]{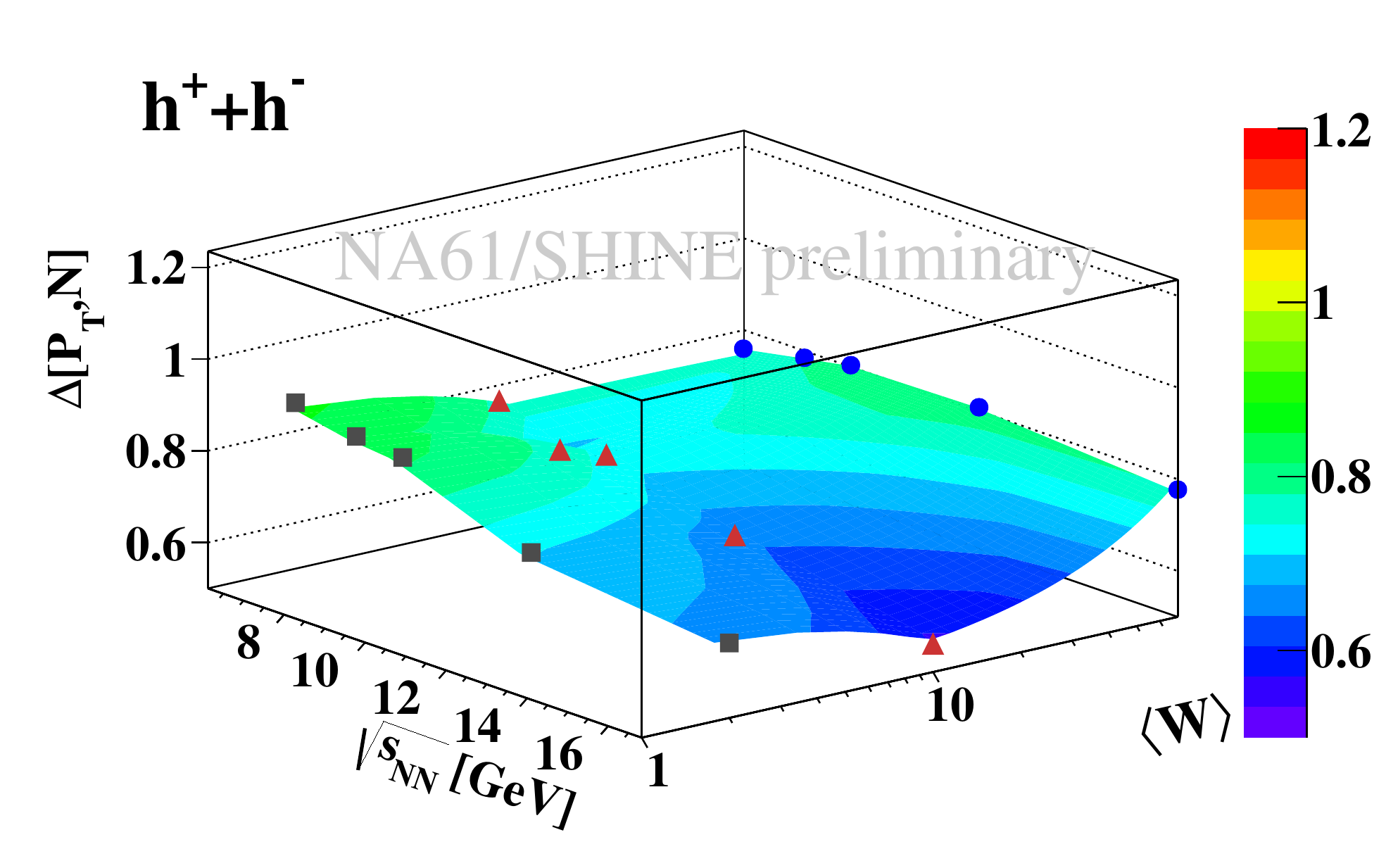}\includegraphics[width=4.5cm]{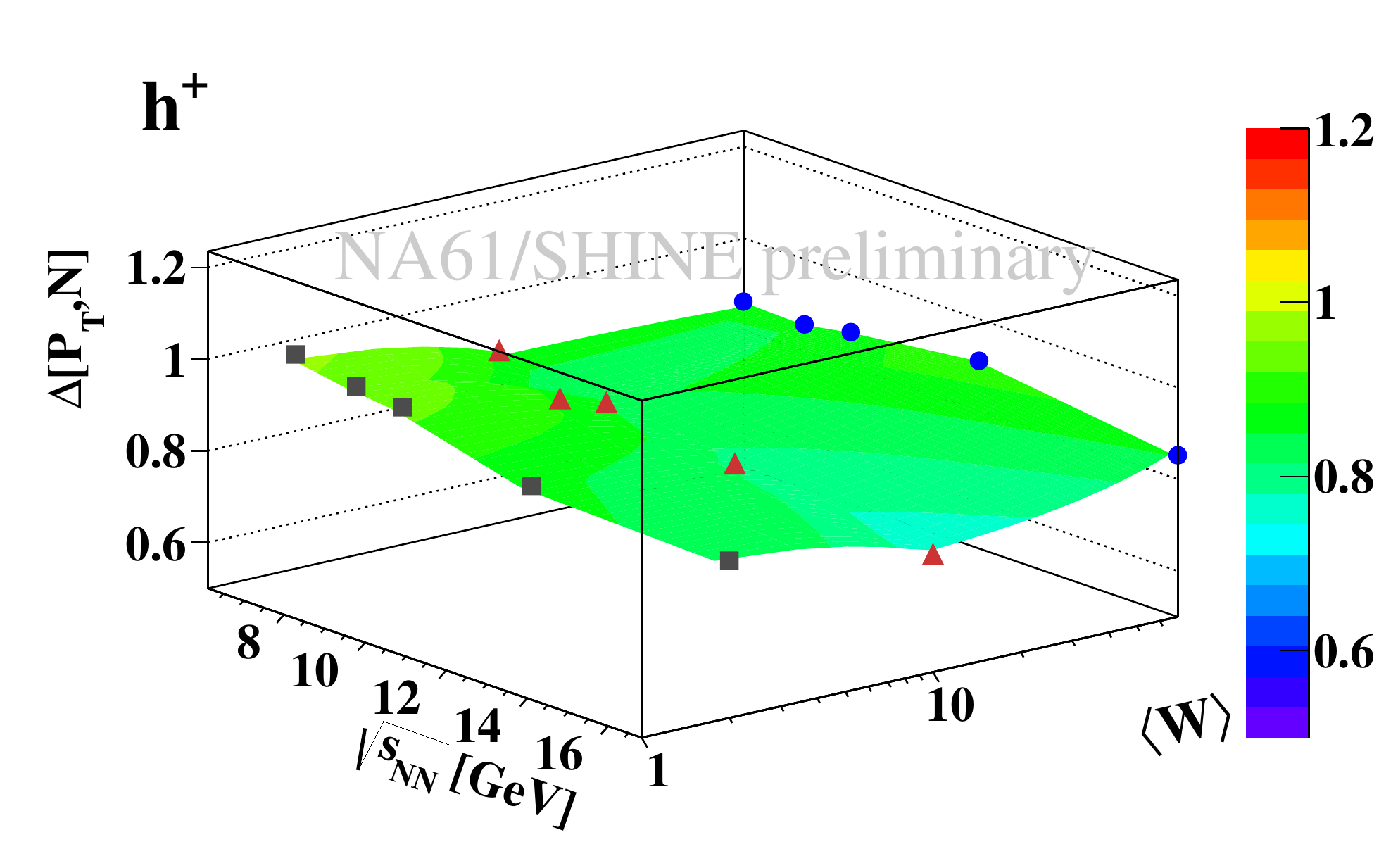}\includegraphics[width=4.5cm]{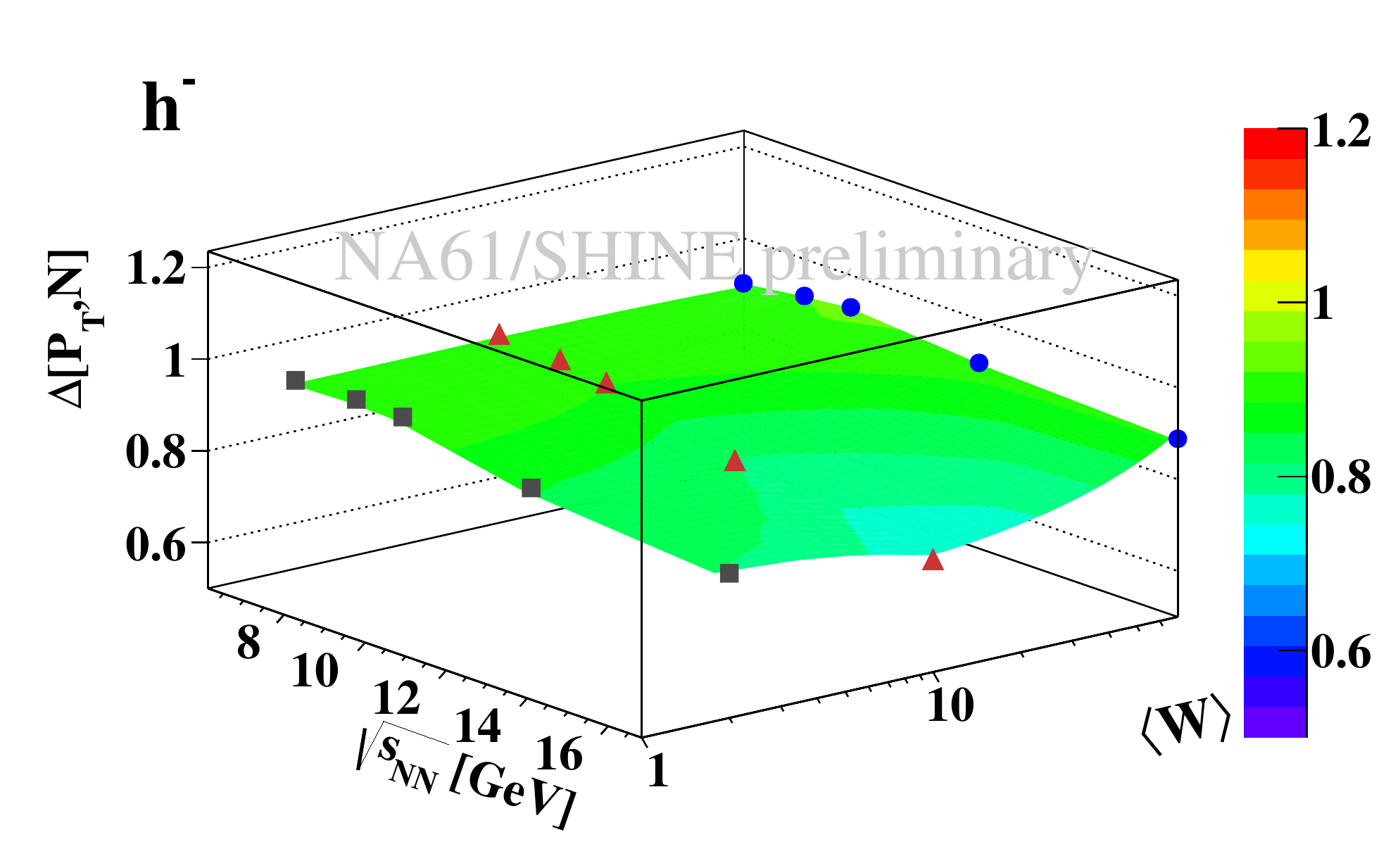}}
\caption{$\Delta[P_T, N]$ as a function of $\sqrt{s_{NN}}$ and $\langle W\rangle$.}
\label{Fig:F2HDelta}
\end{figure}
\begin{figure}[htb]
\centerline{%
\includegraphics[width=4.5cm]{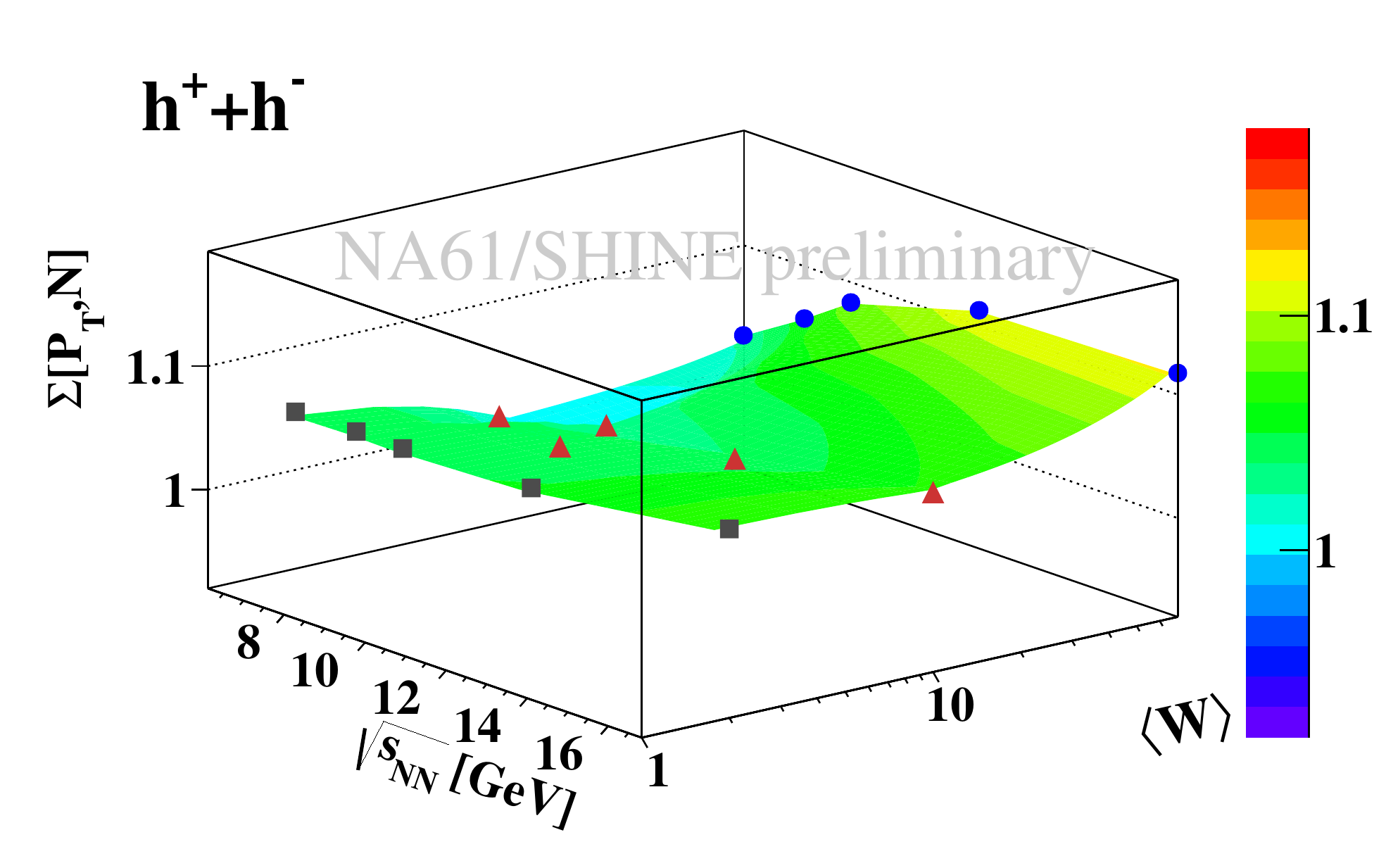}\includegraphics[width=4.5cm]{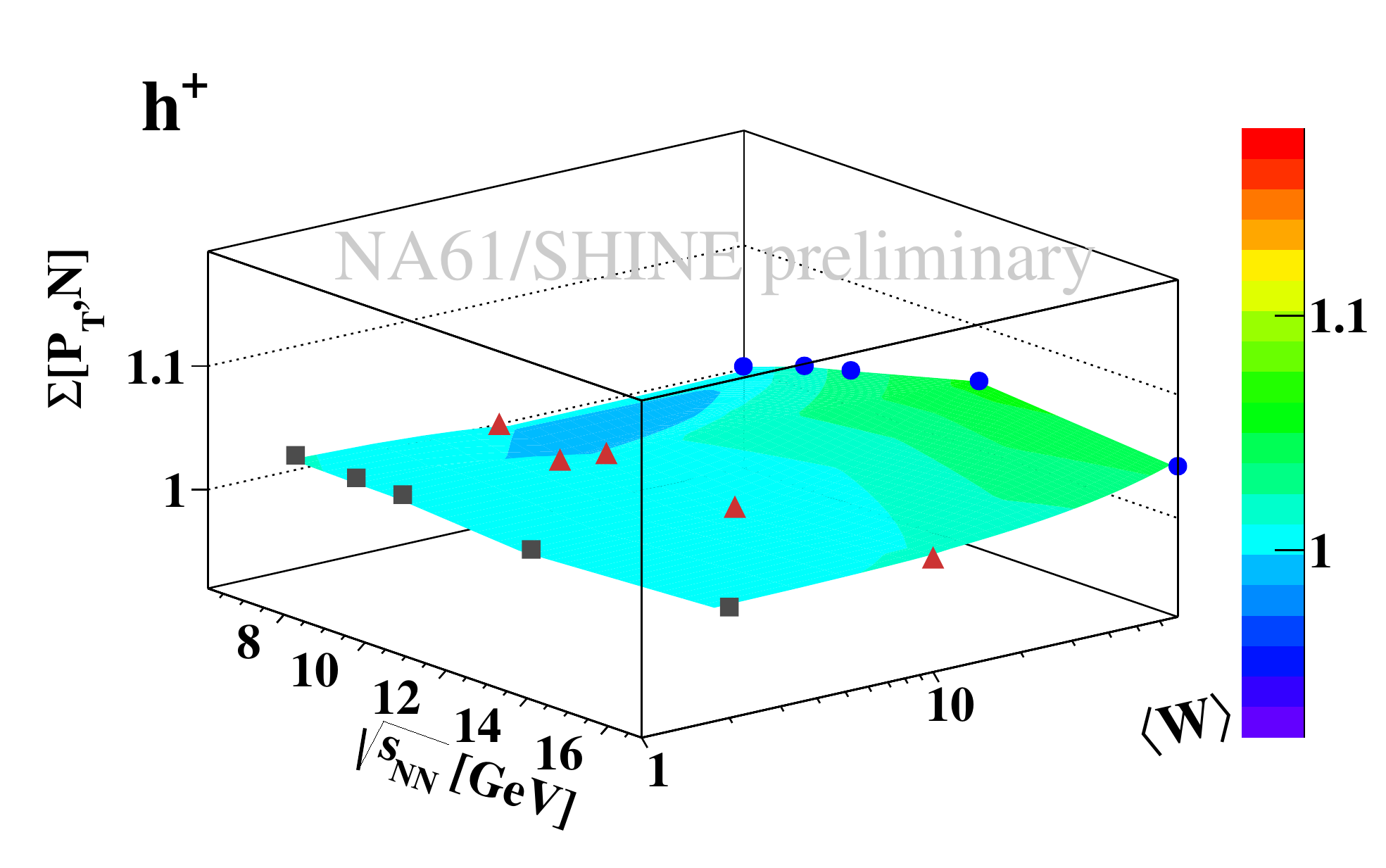}\includegraphics[width=4.5cm]{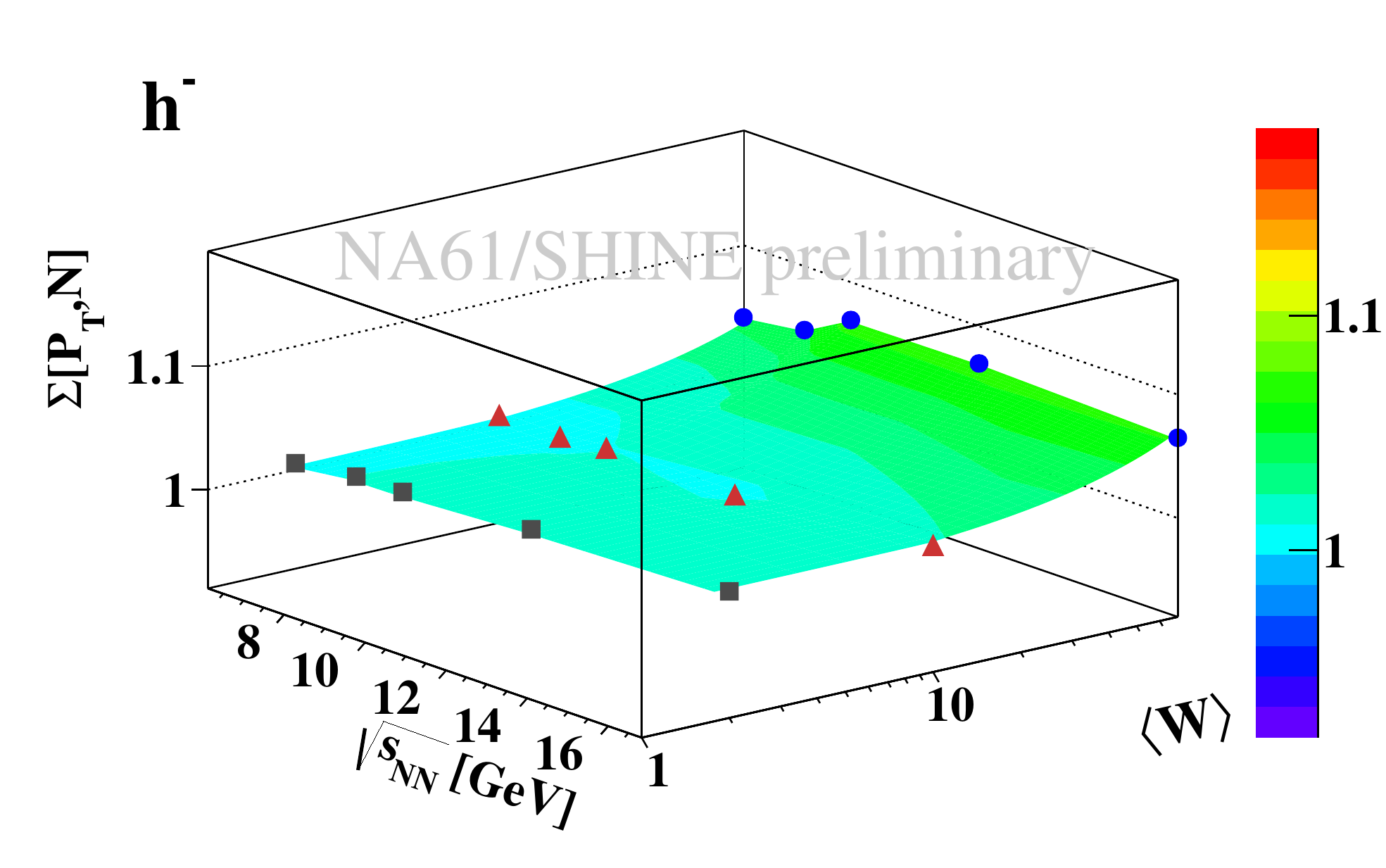}}
\caption{$\Sigma[P_T, N]$ as a function of $\sqrt{s_{NN}}$ and $\langle W\rangle$.}
\label{Fig:F2HSigma}
\end{figure}
\subsection{Comparison with NA49 results}
Measurements done by the NA49 Collaboration for Pb+Pb collisions were performed in a narrower acceptance \cite{NA49:Fluct}. The NA61/SHINE results for central Ar+Sc collisions for all charged hadrons within the NA49 acceptance are shown in Fig. 5 in comparison with the corresponding Pb+Pb results from the NA49 experiment. Both dependences are consistent within the first estimate of the systematic uncertainties for Ar+Sc collisions, suggesting that limitation of acceptance makes fluctuation observables go to a common limit.
\begin{figure}[htb]
\centerline{%
\includegraphics[width=5.5cm]{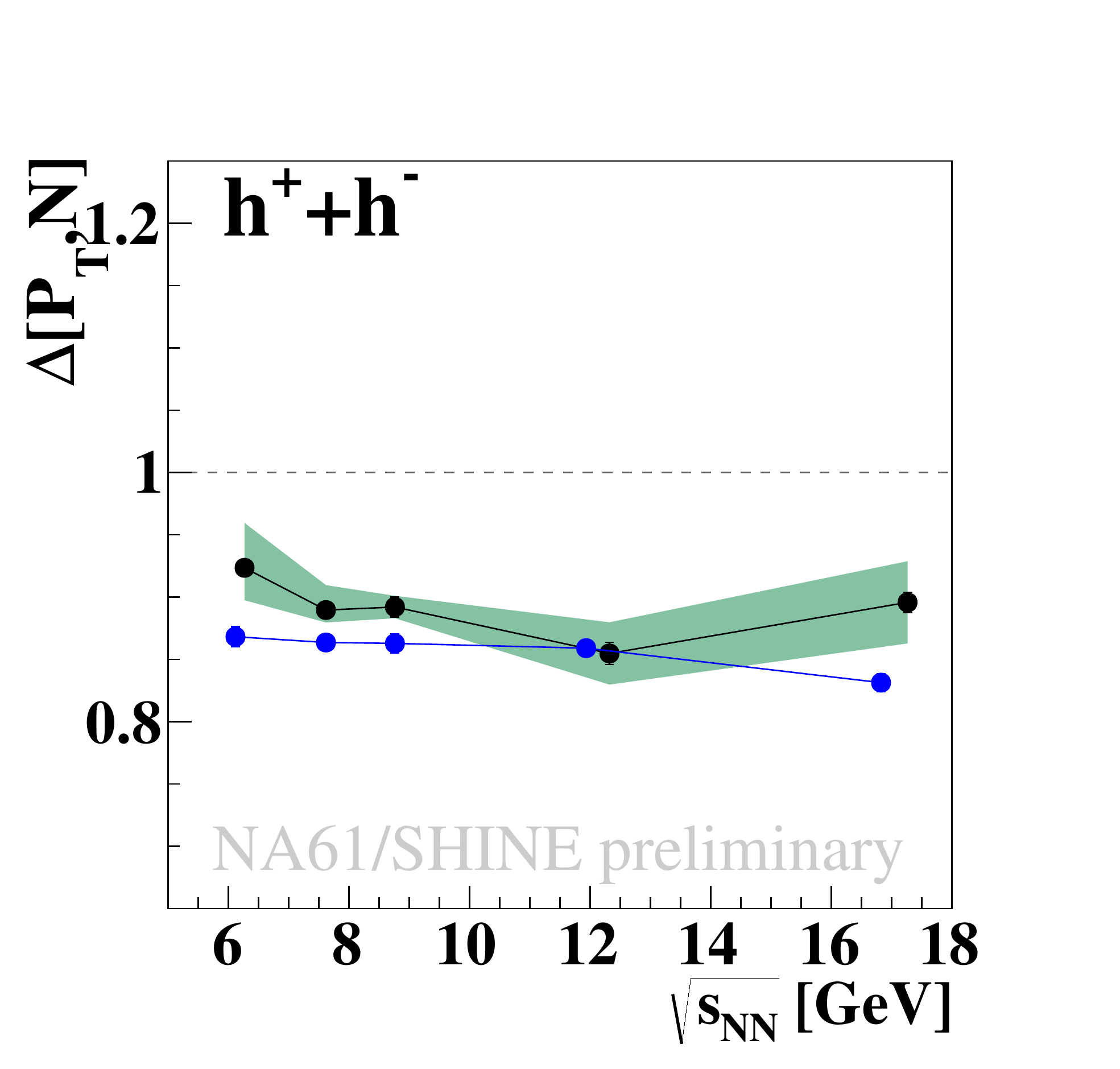}\includegraphics[width=5.5cm]{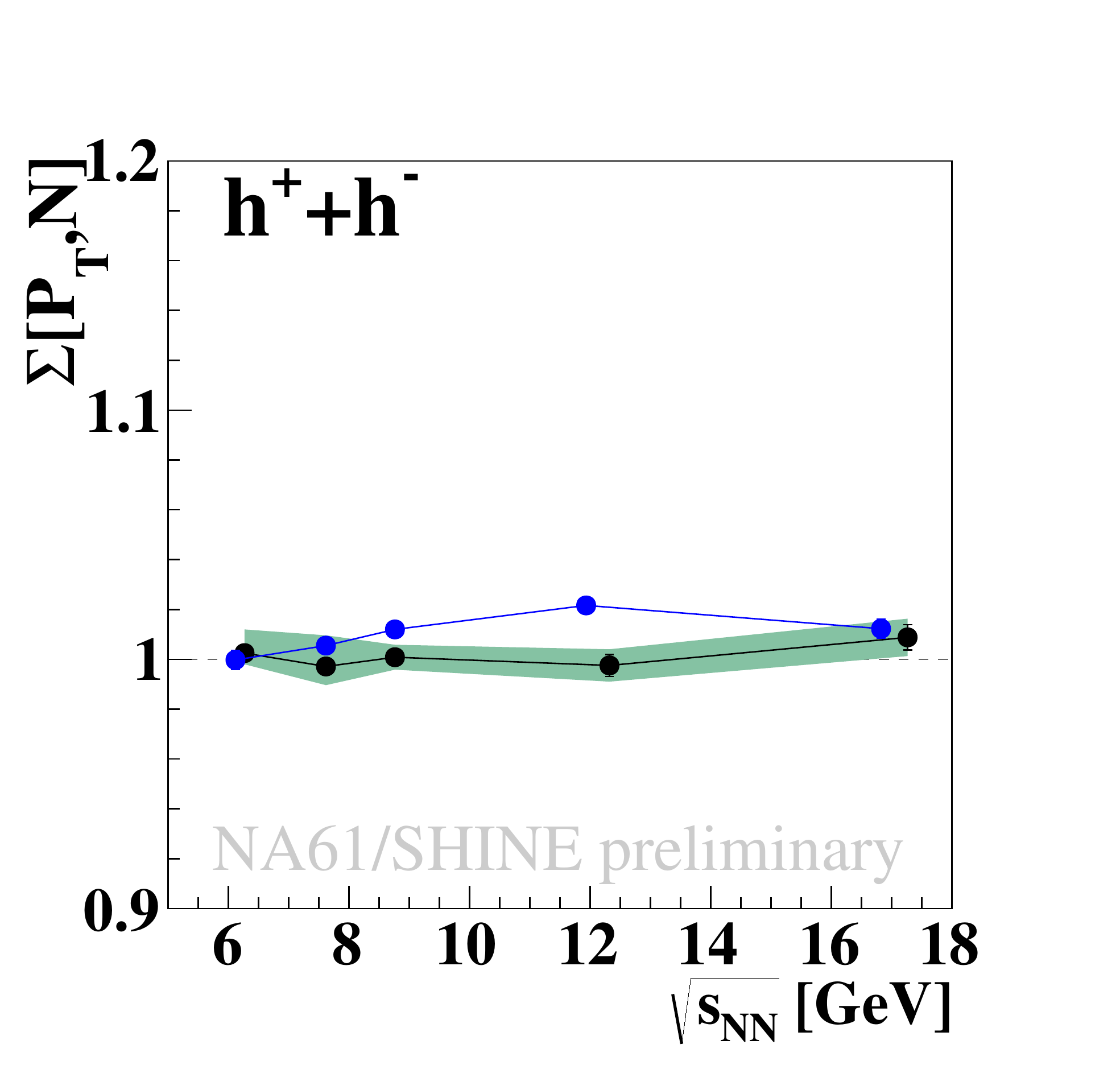}}
\caption{$\Delta[P_T, N]$ and $\Sigma[P_T, N]$ by NA61/SHINE 0-5\% Ar+Sc (blue points) and NA49 0-7.2\% Pb+Pb (black points). Results in $1.1< y_{\pi}< 2.6$ and $y_{p}< y_{beam}-0.5$ with narrow azimuthal acceptance. Systematic uncertainties of the Pb+Pb results are presented with the green bands.}  
\label{Fig:F2HNA49}
\end{figure}
\\

{\small \textbf{Acknowledgments:} This work was supported by the SPbSU research grant 11.38.242.2015.}

\end{document}